\documentclass[fleqn,usenatbib]{mnras}

\usepackage{graphicx}	% Including figure files
\usepackage{amsmath}	% Advanced maths commands
\usepackage{amssymb}	% Extra maths symbols
\usepackage{bm}
\usepackage{newtxtext,newtxmath}
\usepackage[T1]{fontenc}
\usepackage{ae,aecompl}

\DeclareRobustCommand{\VAN}[3]{#2}
\let\VANthebibliography\thebibliography
\def\thebibliography{\DeclareRobustCommand{\VAN}[3]{##3}\VANthebibliography}

\usepackage{multirow}
\newcommand\Cal{CAL\,87}

\usepackage{xcolor}
\usepackage{import}

\title[The period evolution of \Cal]{Revised ephemeris and orbital period derivative of the supersoft X-ray source \Cal\space based on 34 years of observations}

\author[Stecchini et al.]{P. E. Stecchini$^{1,2}$\thanks{E-mail: paulo.stecchini@usp.br / paulo.stecchini@inpe.br},
F. Jablonski$^{2}$,
M. P. Diaz$^{1}$,
F. D'Amico$^{2}$,
A. S. Oliveira$^{3}$,
N. Palivanas$^{3}$ and
\newauthor R. K. Saito$^{4}$\vspace{0.4cm}\\
\parbox{\textwidth}{$^{1}$IAG, Universidade de São Paulo, Rua do Matão 1226, 05508–900,  São Paulo--SP, Brazil\\
$^{2}$Divisão de Astrofísica, Coordenação de Engenharia, Tecnologia e Ciências Espaciais, Instituto Nacional de Pesquisas Espaciais, Av. dos Astronautas 1758, 12227-010, S.J. Campos--SP, Brazil\\
$^{3}$IP\&D, Universidade do Vale do Paraíba, Av. Shishima Hifumi, 2911, 12244-000, S.J. Campos--SP, Brazil\\
$^{4}$Departamento de F\'isica, Universidade Federal de Santa Catarina, Trindade 88040-900, Florian\'opolis--SC, Brazil
}}

\date{Accepted XXX. Received YYY; in original form ZZZ}

\pubyear{2023}

\begin{document}
\label{firstpage}
\pagerange{\pageref{firstpage}--\pageref{lastpage}}
\maketitle

\begin{abstract}
In this study, we present an analysis of over 34 years of observational data from \Cal, an eclipsing supersoft X-ray source. The primary aim of our study, which combines previously analysed measurements as well as unexplored publicly available datasets, is to examine the orbital period evolution of \Cal. After meticulously and consistently determining the eclipse timings, we constructed an O$-$C (observed minus calculated) diagram using a total of 38 data points. Our results provide confirmation of a positive derivative in the system's orbital period, with a determined value of $\dot{P}=+\,8.18\pm1.46\times10^{-11}$\,s/s. We observe a noticeable jitter in the eclipse timings and additionally identify a systematic delay in the X-ray eclipses compared to those observed in longer wavelengths. We discuss the interplay of the pertinent factors that could contribute to a positive period derivative and the inherent variability in the eclipses.
\end{abstract}


%%%%%%%%%%%%%%%%%%%%%%%%%%%%%%%%%%%%%%%%%%%%%%%%%%

%%%%%%%%%%%%%%%%% BODY OF PAPER %%%%%%%%%%%%%%%%%%

\section{Introduction}
\label{sect:01}

Supersoft X-ray sources (SSSs) are characterized by their very soft emission in X-rays (below $\sim$\,1\,keV), blackbody-like spectra of effective temperatures $kT$\,$\sim$\,20--100\,eV and high bolometric luminosities of $\sim$\,10$^{36-38}$\,erg\,s$^{-1}$ \citep[e.g.][]{1997ARA&A..35...69K}. \cite{1992A&A...262...97V} showed that the soft X-ray emission observed in SSSs originates from stable hydrogen burning on the surface of a massive ($\gtrsim$\,1\,M$_{\odot}$) white dwarf (WD) which accretes at very high rates ($\sim$\,10$^{-7}$\,M$_{\odot}$\,year$^{-1}$) from its companion. For this high mass accretion to be sustained, argued \cite{1992A&A...262...97V}, the companion star (with mass $M_2$) of the binary system should be even more massive than the WD ($M_1$), i.e. a mass ratio $M_2/M_1=q>1$, which would lead to dynamically unstable mass transfer via Roche lobe overflow (RLOF). An alternative scenario was proposed by \cite{1998A&A...338..957V}, in which wind-driven mass transfer (WDMT) could also account for the high accretion rates in systems where the companion star is less massive, i.e. $q<1$. This would also explain SSSs with orbital periods shorter than $\sim$\,6\,hr.

\Cal\space is a well known SSS located in the Large Magellanic Cloud (LMC). It was, along with CAL\,83, the first supersoft object detected \citep{1981ApJ...248..925L}, considered thus to be a prototype of the class. By placing observational parameters obtained from X-ray spectral analyses of \Cal\space in a luminosity-temperature diagram, computed for surface hydrogen-burning WDs, \cite{2004ApJ...612L..53S} inferred that the WD in the system might be as massive as 1.35\,M$_{\odot}$.  Optical spectroscopic data revealed Balmer absorpion lines whose radial velocities suggest that -- for a 1.3--1.4\,M$_{\odot}$ compact object -- the donor should have no more than 0.4\,M$_{\odot}$ \citep{1998ApJ...502..408H}. An orbital evolutionary analysis performed by \cite{2007A&A...472L..21O} pointed to a 0.34\,M$_{\odot}$ donor, corroborating the low mass nature of the companion star in \Cal. These values lead to a mass ratio of $q\approx0.25-0.3$. 

Light curve analyses have shown that \Cal\space exhibits partial eclipses in optical, \citep{1989MNRAS.241P..37C, 1990ApJ...350..288C, 1997MNRAS.287..699A, 2007A&A...472L..21O}, ultraviolet \citep[UV,][]{1995AJ....110.2394H}, and X-rays \citep{1993PASP..105..863S, 1998ApJ...503L.143A, 2004RMxAC..20...18G,  2010AN....331..152E} with a period of approximately 0.44\,d (10.6\,h). The depths of the eclipses imply a high orbital inclination \citep[$>$\,70$^{\circ}$, e.g.][]{1997A&A...318...73S}. An increase in the system's orbital period ($\dot{P}$\,=\,$+\,1.7\,\pm\,0.3\,\times\,10^{-10}$\,s/s) was reported by \cite{2007A&A...472L..21O} based on an estimation from two optical minima that were approximately 9000 cycles apart. \cite{2014ApJ...792...20R} also derived a positive value ($\dot{P}$\,=\,$+\,6\,\pm\,2\,\times\,10^{-10}$\,s/s) from the X-ray eclipse phase shift (relative to longer wavelengths) present in \textit{XMM-Newton}'s data of \Cal. The authors of both aforementioned studies have advocated that the increasing orbital period, coupled with the low mass ratio, favours the WDMT model; \cite{2015ApJ...815...17A} demonstrated through analytical and numerical calculations of orbital evolution that the reported dynamical parameters of \Cal\space indeed admit such a scenario.

In this study, we compiled archival data spanning a 34-year time baseline for \Cal. The primary goal of this compilation was the construction of an O$-$C (observed minus calculated) diagram, which serves as the basis for a comprehensive analysis of orbital period variations and other related phenomena. While certain datasets have been previously analysed, our approach includes several previously unexamined light curves, rendering them novel in the context of this investigation.

\section{Data collection and determination of eclipse timings}
\label{sect:02}

The times of minimum were essentially obtained through two different approaches: taken from published ephemerides or determined by us after analysing publicly available data. In the latter case, the data points used to build the light curves are sourced from either previously published measurements or archival data publicly available in survey/project catalogues. 

Eclipse timings seized directly from published works are those from CTIO \citep[one minimum -- B, V and R bands altogether;][]{1990ApJ...350..288C}, ESO \& SAAO \citep[one minimum -- B and V bands altogether;][]{1989MNRAS.241P..37C}, and SOAR \citep[V band;][]{2007A&A...472L..21O}. They are marked with an asterisk in Table\,\ref{tab:01}. The HST (UV band) and a second CTIO (V band) minima were determined based on measurements presented in table 1 of \cite{1995AJ....110.2394H} and table 2 of \cite{1998ApJ...502..408H}, respectively.

All other datasets were collected from public catalogues. These are the light curves from MACHO\footnote{\url{http://cdsarc.u-strasbg.fr/viz-bin/VizieR-5?-ref=VIZ642dda8b315c&-out.add=.&-source=II/247/machovar&recno=4458}} (two minima -- a "blue" and a "red" band), OGLE\footnote{\url{https://ogledb.astrouw.edu.pl/~ogle/OCVS/?OGLE-LMC-ECL-24119}} (two minima, a V and a I band), VMC\footnote{\url{https://archive.eso.org/scienceportal/home?data_collection=VMC}} (one minimum -- Y, J, K${_\text{S}}$ bands altogether),  and TESS\footnote{\url{https://mast.stsci.edu}} (28 minima, in a wide I band). Except for TESS (which will be explained shortly), the aforementioned data were straightforwardly retrieved as magnitudes versus time. Data from MACHO had already been analysed by \cite{1997MNRAS.287..699A}, who reported the ephemeris \hbox{$T_{0,\rm{Alcock}}=\rm{MJD}\,50111.0144(3)+0.44267714(6)$}, most commonly cited up to the present day. 
 
The TESS light curves were produced from aperture photometry on the Full Frame Images (FFIs) of 31 sectors. Each sector spans about 26 days of observations, at a cadence of 30, 10, or 3 minutes. Three of the sectors were discarded due to the target being too close to the edge of the detector. The aperture photometry was performed with Lightkurve \citep{2018ascl.soft12013L}, an open source Python package written for TESS and Kepler data analysis. From each FFI we made a \hbox{11\,$\times$\,11} pixels cutout centred on the target. The light curves were obtained from those cutouts by measuring the total flux in a \hbox{2\,$\times$\,2} pixel mask that includes the target, subtracted of the sky average flux obtained from a background mask. The resulting data are given in flux (e$^-$\,s$^{-1}$) versus Barycentric Tess Julian Date (BTJD), defined as \hbox{BTJD = BJD $-$ 2,457,000,} where BJD is the Barycentric Julian Date.

\begin{figure}
% \centering
	\includegraphics[width=\columnwidth]{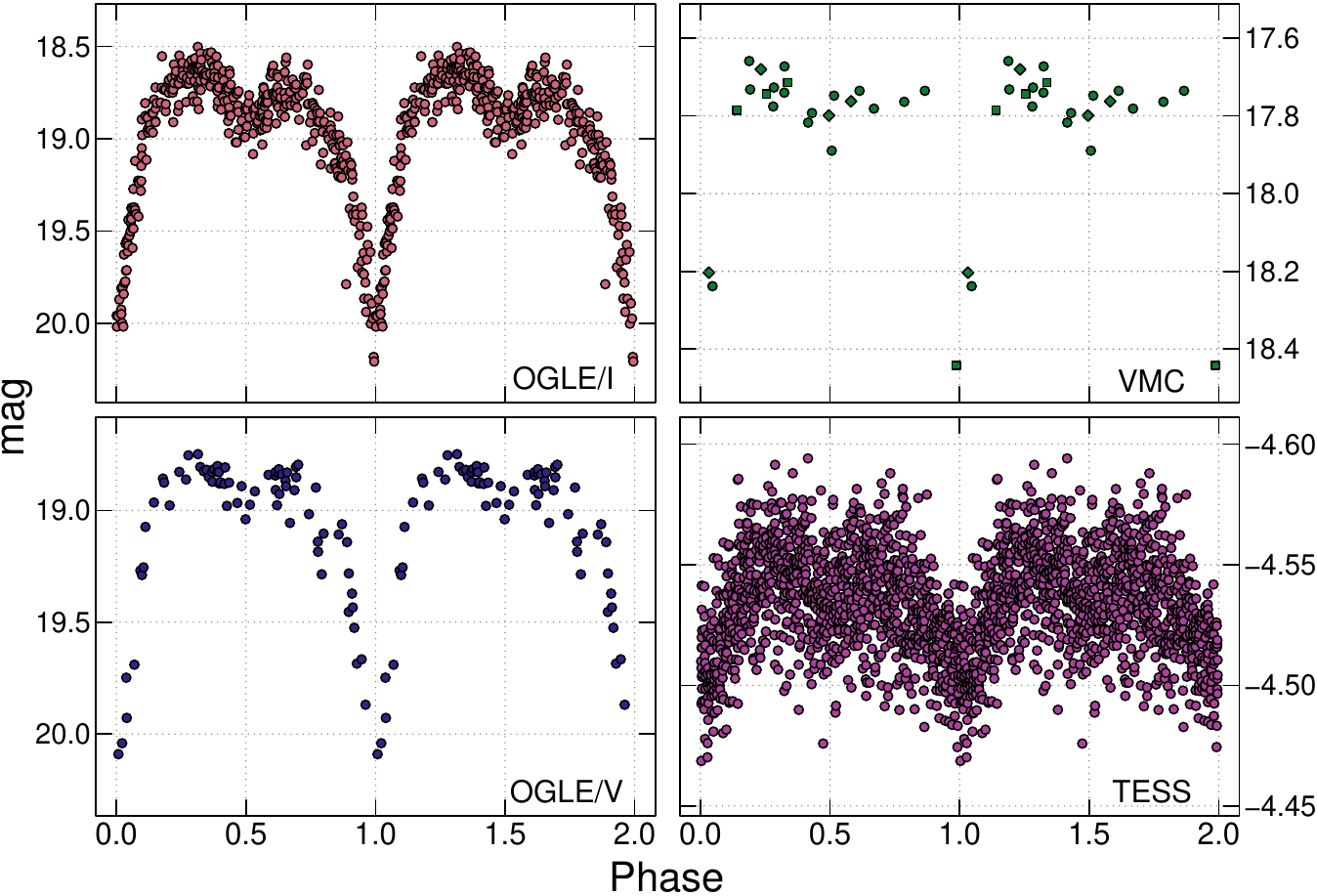}
    \caption{Folded light curves of \Cal\space from archival data not previously analysed. The light curves are folded with the linear ephemeris presented in equation\,\ref{eq:01} (Section\,\ref{sect:03}). Two cycles are shown for clarity. OGLE data (left panels): vertical axes show the actual magnitude measured in bands I (upper) and V (bottom). VMC data (upper right panel): circles, squares and diamonds refer, respectively, to measurements in bands K$_{\text{S}}$, Y and J. Vertical axis shows magnitude for K$_{\text{S}}$; the Y and J data points were shifted to match the K$_{\text{S}}$ level out of eclipse. TESS (sector 11) data (bottom right panel): straightforwardly converted from flux to magnitude ($-$2.5\,log$f$), i.e. the actual units for the vertical axis are $\Delta$mag. The relatively small amplitude and large scatter are due to light dilution caused by the presence of numerous other sources within the TESS 2\,$\times$\,2 pixels extraction aperture where \Cal\space is found. }
    \label{fig:01}
\end{figure}

Some of the data have not been previously mentioned in the literature; this applies to the measurements of \Cal\space from OGLE, VMC and TESS. Their folded light curves are displayed in Fig.\,\ref{fig:01}. For TESS, from which we use data from multiple sectors and thus obtain multiple eclipsing times, we present the light curve from one sector.

While our primary focus was not on incorporating X-ray eclipses into the period evolution analysis, we retrieved historical X-ray light curves to examine their correlation with those at longer wavelengths. X-ray data reduction and light curves production were accomplished following each mission's standard procedures: XSELECT tasks from the FTOOLS package (HEASOFT, v. 6.32) for PSPC data from \textit{ROSAT}, the \textit{Chandra} Interactive Analysis of Observations (CIAO, v. 4.15.0) for ACIS-S data from \textit{Chandra}, and the \textit{XMM--Newton} Science Analysis System (SAS, v. 20.0.0) for EPIC-pn data from \textit{XMM--Newton}. All light curves were extracted for the energy band \hbox{0.2--1\,keV}. \textit{ROSAT}'s observation spanned over 60 hours, although there were numerous gaps. \textit{Chandra} and \textit{XMM-Newton} observations of \Cal\space were continuous over more than one orbital cycle, encompassing 3 and 2 full eclipses, respectively. The phase diagrams for the X-ray data are presented in Fig.\,\ref{fig:03} later in this paper.

To determine a composite time of minimum for each dataset, we started with a provisional epoch close to the middle of the dataset. The precise time of minimum was extracted with respect to this epoch using a spline fit with the aid of the {\tt loess()} function in the {\tt R} statistical package. To estimate the uncertainty in the derived time of minimum, a bootstrap technique was employed. This technique involves resampling each flux value around its nominal value based on the corresponding 1-$\sigma$ error bar. The median of the absolute deviations from one thousand realisations of this procedure was used to calculate the standard deviation of the timing. In Table\,\ref{tab:01} we present the final epochs of minimum and their uncertainties, keeping the cycle number relative to the ephemeris of \citet{1997MNRAS.287..699A}, and general information regarding the observations. For brevity, all the time stamps are expressed in MJD (Modified Julian Date), i.e. HJD or BJD $-$ 2,400,000.5.

% \usepackage{multirow}
\begin{table}

\caption{Epochs of minimum.}
\label{tab:01}
\hspace*{-12pt}
\resizebox{\columnwidth}{!}{
\begin{tabular}{ccccc} %\begin{tabular}{ccccc}
\hline
Epoch of minimum & Uncertainty in epoch & \multirow{2}{*}{\phantom{$^{\dag}$}Cycle$^{\dag}$} & \multirow{2}{*}{Provenance} & \multirow{2}{*}{Band} \\
(MJD)            & (days)               &                        &                             &                       \\ \hline
\phantom{$^*$}47506.30210$^*$     & 0.0002               & -5884                  & CTIO                        & B, V, R               \\
\phantom{$^*$}47531.09520$^*$      & 0.0030               & -5828                  & ESO \& SAAO                 & B, V                  \\
48306.67950      & 0.0110               & -4076                  & ROSAT                       & \phantom{$^{\text{a}}$}X-rays$^{\text{a}}$                \\
49724.11870      & 0.0060               & -874                   & HST                         & \phantom{$^{\text{b}}$}UV$^{\text{b}}$                    \\
50028.23574      & 0.0024               & -187                   & CTIO                        & V                     \\
50180.51280      & 0.0015               & 157                    & MACHO                       & \phantom{$^{\text{c}}$}b$^{\text{c}}$                     \\
50182.72351      & 0.0019               & 162                    & MACHO                       & \phantom{$^{\text{d}}$}r$^{\text{d}}$                     \\
52135.38977      & 0.0079               & 4573                   & Chandra                     & X-rays                \\
52748.49518      & 0.0014               & 5958                   & XMM-Newton                  & X-rays                \\
\phantom{$^*$}53680.32440$^*$      & 0.0003               & 8063                   & SOAR                        & V                     \\
56001.28307      & 0.0021               & 13306                  & OGLE                        & I                     \\
56022.53521      & 0.0011               & 13354                  & OGLE                        & V                     \\
56632.97981      & 0.0060               & 14733                  & VMV                         & Y, J, K${_\text{S}}$  \\
58338.62551      & 0.0039               & 18586                  & TESS                        & \phantom{$^{\text{e}}$}Wide I$^{\text{e}}$                \\
58367.39827      & 0.0019               & 18651                  & TESS                        & Wide I                \\
58423.17168      & 0.0040               & 18777                  & TESS                        & Wide I                \\
58450.17719      & 0.0051               & 18838                  & TESS                        & Wide I                \\
58478.51156      & 0.0053               & 18902                  & TESS                        & Wide I                \\
58503.30069      & 0.0042               & 18958                  & TESS                        & Wide I                \\
58529.42007      & 0.0056               & 19017                  & TESS                        & Wide I                \\
58555.53251      & 0.0029               & 19076                  & TESS                        & Wide I                \\
58582.09385      & 0.0025               & 19136                  & TESS                        & Wide I                \\
58609.99029      & 0.0042               & 19199                  & TESS                        & Wide I                \\
58638.76056      & 0.0042               & 19264                  & TESS                        & Wide I                \\
59047.79394      & 0.0053               & 20188                  & TESS                        & Wide I                \\
59073.91511      & 0.0046               & 20247                  & TESS                        & Wide I                \\
59100.91171      & 0.0046               & 20308                  & TESS                        & Wide I                \\
59156.68956      & 0.0036               & 20434                  & TESS                        & Wide I                \\
59186.79344      & 0.0040               & 20502                  & TESS                        & Wide I                \\
59241.25008      & 0.0034               & 20625                  & TESS                        & Wide I                \\
59267.37093      & 0.0024               & 20684                  & TESS                        & Wide I                \\
59293.48852      & 0.0033               & 20743                  & TESS                        & Wide I                \\
59319.60339      & 0.0043               & 20802                  & TESS                        & Wide I                \\
59346.60482      & 0.0045               & 20863                  & TESS                        & Wide I                \\
59375.37238      & 0.0020               & 20928                  & TESS                        & Wide I                \\
59975.20202      & 0.0019               & 22283                  & TESS                        & Wide I                \\
60000.88383      & 0.0043               & 22341                  & TESS                        & Wide I                \\
60026.99755      & 0.0037               & 22400                  & TESS                        & Wide I                \\
60053.99849      & 0.0016               & 22461                  & TESS                        & Wide I                \\ 
60081.89137      & 0.0054               & 22524                  & TESS                        & Wide I                \\ 
60110.22337      & 0.0056               & 22588                  & TESS                        & Wide I                \\ \hline
\end{tabular}}

$^*$Epochs taken from the literature.

$^{\dag}$Based on the cycle counting of \citet{1997MNRAS.287..699A}.

$^{\text{a}}$0.2--1.0\,keV.
$^{\text{b}}$1350--2200\,\AA.
$^{\text{c}}$$\sim$V.
$^{\text{d}}$$\sim$R$+$I.
$^{\text{e}}$$\sim$6000--10000\,\AA.
\end{table}

\section{The O$-$C diagram for CAL\,87}
\label{sect:03}

The best-fitting weighted linear ephemeris ($T_{\text{min}}\,=\,T_{\text{0}} + P_{\text{0}}E$), calculated when considering all epochs of minimum listed in Table\,\ref{tab:01} -- except for the X-ray data -- is
\begin{equation}
\label{eq:01}
T_{\text{min}}\,=\,\text{MJD}\,50111.01655(28)\,\pm\,0.442677572(33)\,E,
\end{equation}
where $E$ is the number of elapsed cycles since $T_{\text{0}}$.

\begin{figure}
% \centering
	\includegraphics[width=\columnwidth]{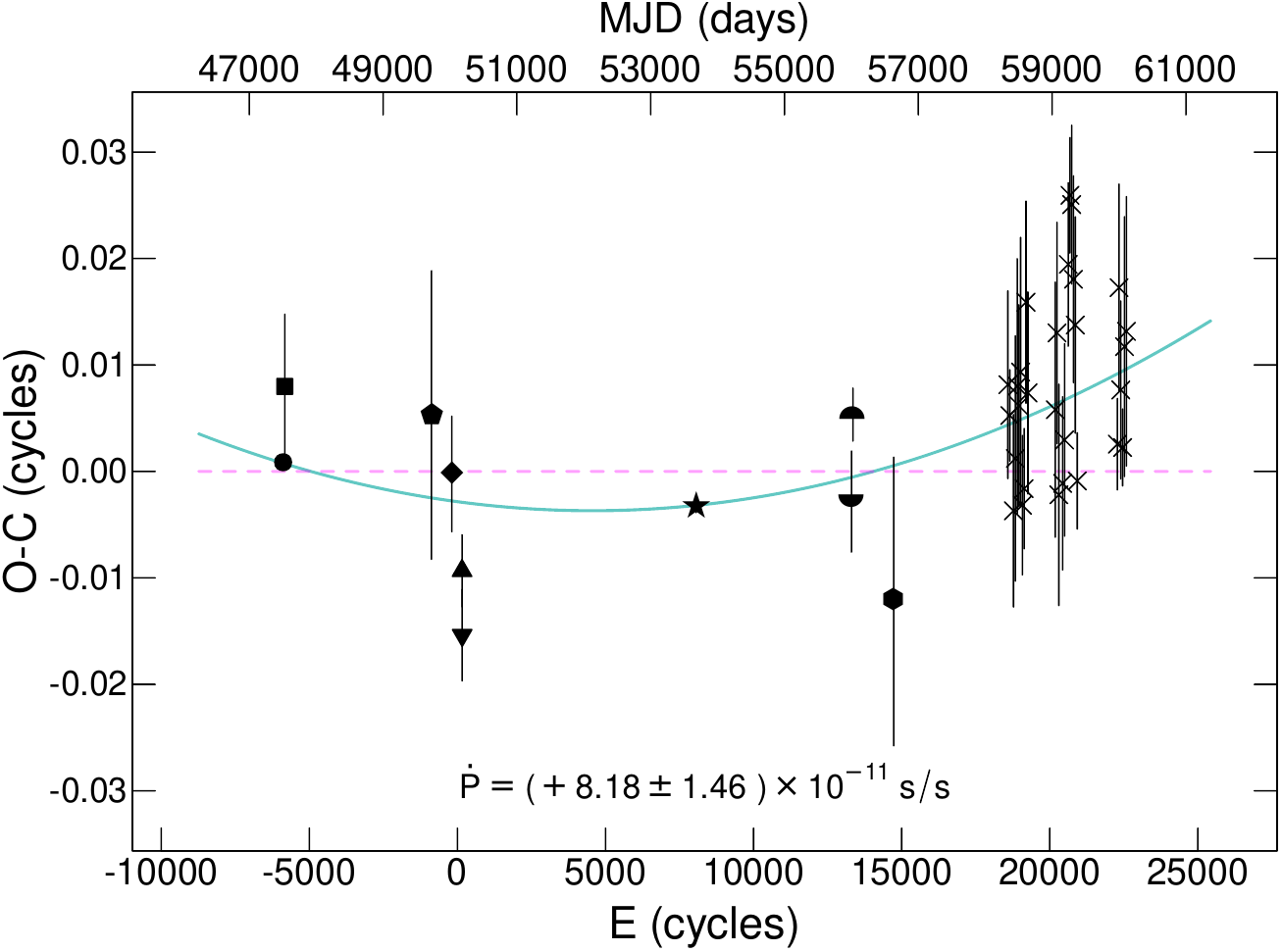}
    \caption{O$-$C diagram computed for the data presented in Table\,\ref{tab:01} (except for the X-ray data). Both computations incorporated the uncertainties of the measurements as weighting factors. Vertical error bars are shown for all data points, although in some cases the bars are smaller than the symbols representing them. Circle: epoch taken from \citet{1990ApJ...350..288C}; square: epoch taken from \citet{1989MNRAS.241P..37C}; pentagon: HST/UV data, taken from \citet{1995AJ....110.2394H}; diamond: CTIO data, taken from \citet{1998ApJ...502..408H}; up-pointing triangle: MACHO/b; down-pointing triangle: MACHO/r; star: epoch taken from \citet{2007A&A...472L..21O}; top half circle: OGLE/I; bottom half circle: OGLE/V; hexagon: VMC; crosses: TESS. Dashed magenta line and solid green line are linear and quadratic fits, respectively. The fits keep the cycle counting of \citet{1997MNRAS.287..699A}.}
    \label{fig:02}
\end{figure}

To investigate a potential orbital period derivative ($\dot{P}$), we incorporated a quadratic term in the ephemeris using the well-known Taylor expansion (i.e. $T_{\text{min}} = T_{\text{0}} + P_{\text{0}}E + \frac{1}{2}P_{\text{0}}\dot{P}E^2$, keeping the same notation). The resulting weighted parabolic fit yields
\begin{multline}
\label{eq:02}
T_{\text{min}}\,=\,\text{MJD}\,50111.01530(13)\,+\,0.442677407(34)\,E\,+\\+1.81(32)\,\times\,10^{-11}\,E^2,
\end{multline}
which translates to a time derivative of \hbox{$\dot{P}=+8.18 \pm 1.46 \times 10^{-11}$\,s/s}. Employing the reduced chi-squared ($\chi^2_{\rm{red}} = \chi^2/\nu$, where $\chi^2$ denotes the standard chi-square statistic and $\nu$ represents the degrees of freedom) as a measure of goodness of fit, we obtain \hbox{$\chi^2_{\rm{red}} = \frac{49.49}{36} = 1.37$} and \hbox{$\chi^2_{\rm{red}} = \frac{37.46}{35} = 1.07$} for the linear and quadratic fits, respectively. Both an F-test and a $\chi^2$ difference test show that there is less than a 0.1\% probability that the null hypothesis (i.e. the linear model) is the best choice. Moreover, upon computing the Akaike Weight \citep{aic}, which expresses the likelihood of a model best representing the data among a set of models, we find that the parabolic model is approximately 99\% more plausible in describing the evolution of \Cal's epochs of minimum. The data, along with the linear and quadratic fits, are presented in Fig.\,\ref{fig:02} in the form of an O$-$C diagram.

\begin{figure}
% \centering
	\includegraphics[width=\columnwidth]{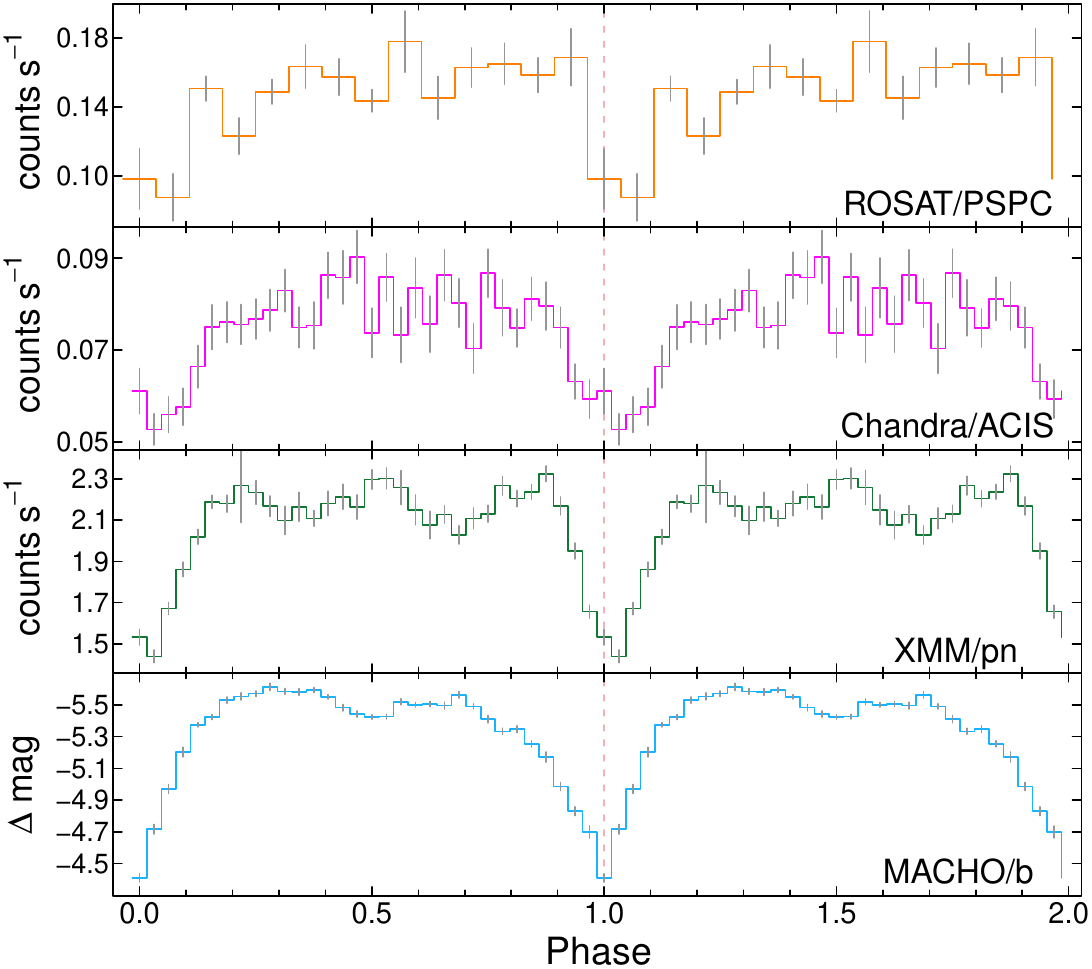}
    \caption{Phase diagrams of \Cal\space to evince the displacement of the X-ray (0.2--1\,keV; first, second and third panels) eclipse minimum time relative to longer wavelengths (optical; bottom panel). Light curves were folded according to the linear ephemeris presented in equation\,(\ref{eq:01}). Two cycles are shown for clarity. Binning consists of 32 bins per phase, except for \textit{ROSAT} (first panel), in which only 14 bins per phase were used due to the uneven distribution of data throughout an orbital cycle.}
    \label{fig:03}
\end{figure}

We have excluded X-ray eclipsing times from the O$-$C calculation based on a previous hint that they may not precisely coincide with the optical eclipse. \cite{2010AN....331..152E}, in their analysis of the \textit{XMM-Newton} observation of \Cal\space (which we also employ here), reported a $\sim$\,0.03\,$\phi$ delay compared to the ephemeris of \cite{1997MNRAS.287..699A}. Indeed, the weighted average of the residuals for the three X-ray minima analysed here (see Table\,\ref{tab:01}), relative to those predicted by equation\,\ref{eq:02}, is 0.02\,$\pm$\,0.01. For clarity and visualization of such a delay, we present in Fig.\,\ref{fig:03} the phase diagram of the three X-ray (\hbox{0.2--1 keV}) light curves (\textit{ROSAT}/PSPC, \textit{Chandra}/ACIS, and \textit{XMM-Newton}/pn), as well as that of MACHO/b, i.e. in the optical band. They are folded according to the linear ephemeris of equation\,(\ref{eq:01}). The observation gaps in \textit{ROSAT}'s data prevented sufficient coverage of all orbital phases; as a result, its folded light curve has a coarser binning (14 bins per phase) compared to the other three, which have 32 bins per phase. We have not attempted to derive an ephemeris for the X-ray eclipses alone due to i) the small number of data points to fit a parabola, ii) the minima's relatively large uncertainties, and iii) their irregular distribution along the time baseline (\textit{XMM-Newton} and \textit{Chandra} timings too close and \textit{ROSAT} timing about 10000 cycles apart). 

\begin{figure}
% \centering
	\includegraphics[width=\columnwidth]{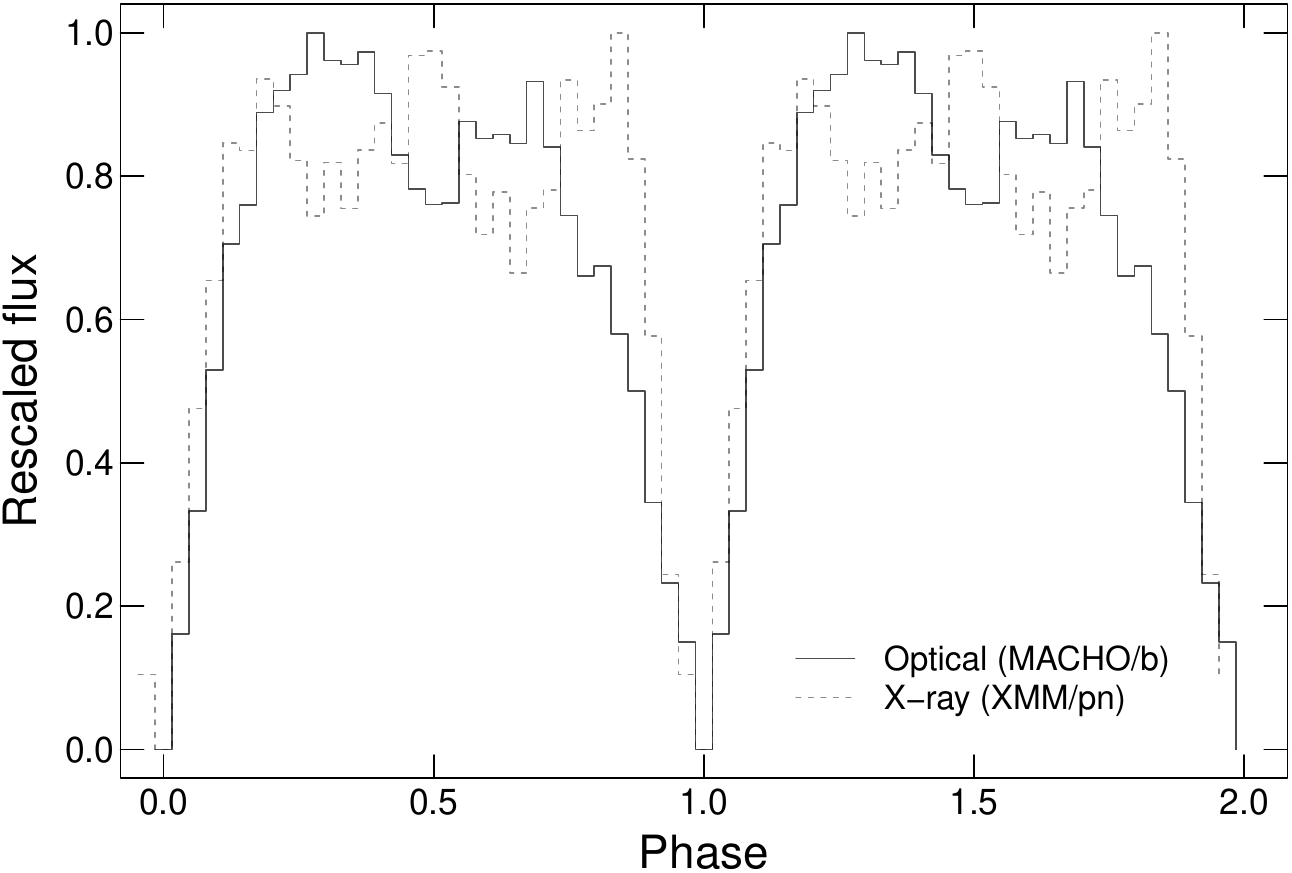}
    \caption{Optical (black solid line) and X-ray (dashed grey line) phase diagrams of CAL\,87, superimposed to highlight shape differences. For comparison purposes, the optical magnitudes have been converted to flux, and the flux values of both light curves have been rescaled to share the same minimum and maximum values. Additionally, the X-ray minimum, originally delayed with respect to the optical (Fig.\,\ref{fig:03}), has been adjusted to match the optical minimum.}
    \label{fig:04}
\end{figure}

At last, to facilitate assessment of the differences in shape of the light curves of CAL\,87 in X-rays and optical, we present again the phase diagrams of \textit{XMM-Newton} and MACHO superimposed in Fig.\,\ref{fig:04}. For a clearer comparison, MACHO magnitudes have been converted to flux by simply calculating 10$^{(-0.4\,\times\,\text{mag})}$, and the flux values of both have then been rescaled to ensure the same range between minimum and maximum. Also, the X-ray minimum has been adjusted to align in phase with the optical minimum.

\section{Discussion and Conclusions}
\label{sect:04}

Conspicuous differences between the eclipse profiles in the optical/IR and X-rays are found. The eclipse seems to display opposite asymmetries: its ingress is prolonged in the optical/IR and steeper in X-rays, whereas the egress is slower than the ingress in X-rays (Fig.\,\ref{fig:04}). In addition to the profiles, the X-ray minimum is observed to be delayed compared to the optical (Fig.\,\ref{fig:03}). These differences indicate that the corresponding sources are not equally distributed around the WD. For instance, if the major contribution to X-rays is associated with the disk-WD boundary layer then the true geometrical conjunction is around phase 0.02. If that is the case, such a change in absolute phasing should be taken into account when deriving binary dynamical solutions from eclipse-phased radial velocity measurements. 
On the other hand, the eclipse timings used in the optical O$-$C diagram (Table\,\ref{tab:01}) have well defined uncertainties. When phased, they exhibit a notable phase jitter (Fig.\,\ref{fig:02}), suggesting an inherently variable phase of minimum light, by a few degrees. One may conjecture that variable asymmetries in the disk's optical emission or in an optically thick wind might explain these small shifts in minimum phase \citep[e.g.][]{1997A&A...321..245M}.  
An anti-correlation between the optical and X-ray flux around phase 0.5 becomes apparent upon comparing the phase-folded curves in Fig.\,\ref{fig:03}, specially from MACHO and \textit{XMM-Newton} data (e.g. Pearson correlation coefficient of --0.59 within phases 0.35 and 0.65; see also Fig.\,\ref{fig:04}), which could be related to the contrasting eclipse asymmetry. Such anti-correlation is less clear for \textit{ROSAT} or \textit{Chandra} light curves, perhaps due to poorer photon statistics. Additional simultaneous multiwavelength observations are required to confirm and quantify this behavior, which has also been seen in other SSSs (e.g. CAL\,83, \citealp{2023MNRAS.522.3472S}).

In order to evaluate the secular binary period evolution it is important to quantify the period change using the most extensive time baseline available. Low-frequency noise in O$-$C diagrams of accreting systems can mislead the observer towards a spurious, yet statistically significant, $\dot{P}$ value.

The observed long-term positive $\dot{P}$ for this system serves as a probe to elucidate the roles of winds, angular momentum losses, and mass ratio in SSSs. Accretion onto the white dwarf may occur via Roche lobe overflow (RLOF) or via wind from a detached irradiated companion. In the context of non-conservative mass transfer, a well-established dynamical relation exists between $\dot{P}/P$, the mass loss rate, $M_{1}$, $M_{2}$, the primary accretion rate, and, for long-period systems, the efficiency of magnetic braking, in the case of a circular orbit (e.g. \citealp{Hilditch2001}). Radiation driven winds are expected to emerge from both the luminous WD's photosphere and the irradiated companion. These winds carry both mass and angular momentum, with the former contributing positively to $\dot{P}$.  Magnetic braking, even in a weak wind from the companion star, introduces a negative contribution. The process of accretion onto the WD itself plays a major role, resulting in either a negative or positive term depending on whether the donor is more massive or less massive than the WD. Our determination of the period derivative favours the hypothesis of mass loss and/or a lighter donor relative to the accreting object. A positive $\dot{P}$ was previously identified for \Cal\space by \citet{2007A&A...472L..21O}, who suggested that the system is powered by the companion wind (the WDMT model) instead of RLOF. \citet{2014ApJ...792...20R} also reported a positive period derivative, which was larger than our value, using a small number of X-ray eclipse timings.

Curiously, the SSS-like binary V\,Sge \citep{1998PASP..110..276S}, which has a comparably long orbital period, seems to exhibit a distinct mass transfer/wind regime. A decreasing orbital period has been claimed for this eclipsing system \citep{1998PASP..110..380P,2022MNRAS.511..553Z}, suggesting a RLOF, high mass transfer rate from the donor. WX\,Cen, another binary system of the V\,Sge class \citep{1995AJ....110.1816D, 2004MNRAS.351..685O}, also shows an orbital period with a negative time derivative \citep{2023ApJ...944...97Z}. In this case, the secondary component is likely less massive than the WD, and thus angular momentum loss driven by a magnetic wind has been proposed as the mechanism for the orbital evolution. In contrast, V617\,Sgr, also known as a V\,Sge-type eclipsing system, exhibits a positive $\dot{P}$ \citep{2006A&A...447L...1S,2023NewA..10302054Z} with the same order of magnitude as found for CAL\,87. A positive $\dot{P}$ is also observed in the long-period eclipsing SSS QR\,And \citep{2023MNRAS.522.2732Z}, which presents an optical orbital light curve remarkably similar to that of \Cal.

Revised dynamical constraints on stellar masses, well-defined long-term period derivatives, and limits on system mass loss are essential for a more accurate understanding of the the mass transfer modes in SSSs. These fundamental quantities remain uncertain or unknown for nearly all bona fide SSSs.
% The last numbered section should briefly summarise what has been done, and describe
% the final conclusions which the authors draw from their work.

\section*{Acknowledgements}

PES acknowledges Conselho Nacional de Desenvolvimento Científico e Tecnológico (CNPq) for financial support under PCI/INPE grant \hbox{\#317428/2023-3}. MPD thanks support from CNPq under grant \hbox{\#305033}. ASO acknowledges São Paulo Research Foundation (FAPESP) for financial support under grant \hbox{\#2017/20309-7}. NP thanks support from Coordenação de Aperfeiçoamento de Pessoal de Nível Superior (CAPES) under grant \hbox{\#88887.823264/2023-00}. R.K.S. acknowledges support from CNPq through projects \hbox{\#308298/2022-5} and \hbox{\#350104/2022-0}. PES, FJ and FD also thank Agência Espacial Brasileira (AEB). The authors thank an anonymous referee for her/his helpful comments and suggestions.

This research is partially based on observations taken within the ESO Public Survey VMC, Programme ID 179.B-2003. This paper also includes data collected by the TESS mission, which are publicly available from the Mikulski Archive for Space Telescopes (MAST). Funding for the TESS mission is provided by the NASA's Science Mission Directorate. This research made use of Lightkurve, a Python package for Kepler and TESS data analysis \citep{2018ascl.soft12013L}. 

%%%%%%%%%%%%%%%%%%%%%%%%%%%%%%%%%%%%%%%%%%%%%%%%%%
\section*{Data Availability}

The data underlying this article are public; the references and/or paths to access are provided in the main text.
% The inclusion of a Data Availability Statement is a requirement for articles published in MNRAS. Data Availability Statements provide a standardised format for readers to understand the availability of data underlying the research results described in the article. The statement may refer to original data generated in the course of the study or to third-party data analysed in the article. The statement should describe and provide means of access, where possible, by linking to the data or providing the required accession numbers for the relevant databases or DOIs.



% \bibliographystyle{mnras}
% \bibliography{example} % if your bibtex file is called example.bib

%\begin{thebibliography}{99}
%\end{thebibliography}

%%%%%%%%%%%%%%%%%%%%%%%%%%%%%%%%%%%%%%%%%%%%%%%%%%

%%%%%%%%%%%%%%%%% APPENDICES %%%%%%%%%%%%%%%%%%%%%

% \appendix

% \section{Some extra material}

% If you want to present additional material which would interrupt the flow of the main paper,
% it can be placed in an Appendix which appears after the list of references.

%%%%%%%%%%%%%%%%%%%%%%%%%%%%%%%%%%%%%%%%%%%%%%%%%%

% Don't change these lines
\bsp	% typesetting comment
\label{lastpage}
\end{document}